\documentclass[12pt,english,floatfix,superscriptaddress,aps,prd,preprint,showkeys,nofootinbib]{revtex4-1}
\usepackage{amsmath}
\usepackage{amssymb}
\usepackage{amsbsy}
\usepackage{amsfonts}
\usepackage{amsopn}
\usepackage{amstext}
\usepackage{graphicx}
\usepackage[english]{babel}
\usepackage{color}
\usepackage{slashed}
\usepackage{esint}
\usepackage[dvips]{epsfig}
\usepackage[dvips]{graphicx}
\usepackage{float}
\usepackage{units}
\usepackage{textcomp}
\usepackage{wasysym}
\usepackage{hyperref}
\usepackage{slashed}
\begin{document}

\title{Probabilistic configurations for thick braneworld in modified symmetric teleparallel gravity}

\author{A. R. P. Moreira}
\email{allan.moreira@fisica.ufc.br}
\affiliation{Reserach Center for Quantum Physics, Huzhou University, Huzhou, 313000, P. R. China.}
\affiliation{Secretaria da Educaç\~{a}o do Cear\'{a} (SEDUC), Coordenadoria Regional de Desenvolvimento da Educaç\~{a}o (CREDE 9),  Horizonte, Cear\'{a}, 62880-384, Brazil.}
\author{Shi-Hai Dong}
\email{dongsh2@yahoo.com}
\affiliation{Reserach Center for Quantum Physics, Huzhou University, Huzhou, 313000, P. R. China.}
\affiliation{Centro de Investigaci\'{o}n en Computaci\'{o}n, Instituto Polit\'{e}cnico Nacional, UPALM, CDMX 07700, Mexico.}

\begin{abstract}
In this research, we delve into the localization patterns of fermionic fields within a braneworld setting, employing a modified gravity model denoted as $f(Q)$. Our investigation revolves around two specific models, $f_1(Q)=Q+kQ^n$ and $f_2(Q)=Q+k_1Q^2+k_2Q^3$
, where we systematically vary the parameters $n$ and $k_{1,2}$.
Through an in-depth analysis encompassing the effective potential, massless, and massive modes, we elucidate how deviations from the conventional symmetric teleparallel equivalent of general relativity (STEGR) gravity impact the localization of fermionic fields. To ensure greater precision, our methodology integrates probabilistic measures such as Shannon entropy and relative probability.
Moreover, we gauge the stability of these models employing differential configurational entropy (DCE), revealing a compelling correlation between the most stable configurations and the emergence of novel structures within the background scalar field.
This work significantly contributes to our understanding of the gravitational modifications' intricate influence on fermionic field localization within braneworld scenarios. By shedding light on these dynamics, it advances the broader comprehension of the interplay between gravity modifications and fermionic field behaviors in these theoretical frameworks.
\end{abstract}

\maketitle

\section{Introduction}
The theory of general relativity has achieved remarkable success in explaining gravitational phenomena in astrophysical contexts involving both weak and strong gravity. However, it faces challenges in addressing phenomena on larger cosmic scales and in providing a comprehensive description of enigmatic physical entities inferred from observational data, such as dark matter and dark energy \cite{Boehm:2000gq,SupernovaSearchTeam:1998fmf}. These unresolved issues have fueled a growing interest in alternative theories of gravity.

Among the myriad of approaches being explored, several promising theories stand out. These include theories that introduce new dynamic degrees of freedom, such as scalar and vector fields \cite{Capozziello:2011et}, as well as theories based on the concept of massive gravitons \cite{Hinterbichler:2011tt}. Additionally, ideas inspired by string theory, such as braneworld models \cite{Maartens:2010ar}, have gained traction. Another innovative direction involves geometric frameworks that diverge from the traditional Riemannian geometry underpinning general relativity.

These diverse theoretical endeavors reflect the vibrant and ongoing quest in the physics community to extend our understanding of gravity beyond the scope of general relativity, aiming to reconcile it with the anomalies and gaps revealed by the latest astronomical observations.

In the realm of alternative gravitational theories, Einstein-Cartan geometry \cite{Hehl:1976kj} and metric-related models, such as $f(R)$ theories \cite{DeFelice:2010aj}, have garnered considerable attention as potential alternatives to general relativity. Another intriguing avenue is the teleparallel equivalent of general relativity (TEGR), conceptualizing gravity as arising from the torsion of spacetime rather than curvature \cite{Aldrovandi}. TEGR employs the \textit{vielbein} field as a dynamic variable, supposing the absence of the Riemann curvature tensor \cite{Aldrovandi}.

More recently, the STEGR has emerged as a particularly compelling alternative. In STEGR, the dynamics of gravitational degrees of freedom involve the non-metricity tensor \cite{Nester:1998mp}. Unlike TEGR, STEGR symmetrically incorporates the metric tensor into its framework. Variants like the $f(Q)$ gravity model have been proposed, offering increased degrees of freedom compared to general relativity, contingent upon coefficients within the Lagrangian \cite{Hohmann:2018wxu,Soudi:2018dhv,BeltranJimenez:2017tkd,BeltranJimenez:2019esp}. The $f(Q)$ gravity model has exhibited notable advancements in characterizing the properties of dark matter and dark energy \cite{Bhar:2023xku,Mussatayeva:2023aoa}. Additionally, substantial progress has been made in studying cosmological phenomena, black holes, and wormholes within this framework \cite{Bajardi:2020fxh, Capozziello:2022tvvi,Capozziello:2022wgl,BeltranJimenez:2019tme,Bhar:2023zwi,Atayde:2023aoj,Koussour:2023rly,Bajardi:2023vcc,Lin:2021uqa,Mustafa:2021ykn}, notably distinguishing it from gravity $f(R)$ theories \cite{BeltranJimenez:2018vdo}.

In the landscape of theoretical physics, braneworld models like the Randall-Sundrum model present innovative approaches to addressing high-energy physics challenges, such as the hierarchy problem \cite{rs,rs2}, and the cosmological constant problem \cite{cosmologicalconstant}. These models have sparked interest in exploring new brane configurations, notably thick branes \cite{DeWolfe:1999cp, Csakil,Gremm2000, Dzhunushaliev:2009}. A distinctive feature of these models is the confinement of standard model matter fields to the brane, while allowing gravity to permeate through extra dimensions. This characteristic has generated significant interest in understanding the mechanisms behind matter field localization on the brane \cite{Gremm1999,CastilloFelisola2004,Navarro2004,BarbosaCendejas2005,Bazeia2007,Liu2011wi,fR1,fR2,tensorperturbations,ftnoncanonicalscalar,ftborninfeld,ftmimetic,Belchior:2023xgn, Moreira:2023uys,Moreira:2023pes,Belchior:2023gmr}.

Particularly intriguing is the investigation into the localization of fermions on the brane, as it presents potential pathways for experimental verification of extra dimensions. This has led to a wealth of research on the topic \cite{RandjbarDaemi2000,Li:2017dkw,Guo:2019vvm,Moreira:2021wkj,Silva:2022pfd,Liu2008,Liu2008b,Liu2009,Liu2009a,Liu2009b,Dantas2013,Moreira20211,Moreira:2023byr,Guerrero:2019qqj,Xie:2019jkq}. Various mechanisms for fermion localization have been proposed, with the simplest and most physically interpretable being the conventional Yukawa coupling between fermion fields and background scalar fields. Recent developments have introduced new mechanisms, such as non-minimal coupling between fermions and curvature \cite{Li:2017dkw,Guo:2019vvm}, and even torsion \cite{Moreira:2021wkj}. To our knowledge, this is the first study investigating fermion localization on the brane within a gravity $f(Q)$ framework, employing Yukawa-type coupling. We also utilize mathematical tools like information entropy and relative probability to probe gravitational influences on fermionic field localization on the brane. Additionally, we conduct a comprehensive analysis of the most probable and stable configurations in $f(Q)$ gravitational models.

Our work is structured as follows: Section \ref{sec1} provides a foundational overview of constructing a gravity model $f(Q)$, and examines two specific models: $f_1(Q)=Q+kQ^n$ and $f_2(Q)=Q+k_1Q^2 +k_2Q^3$, including the energy conditions and solutions for the background scalar field. Section \ref{sec2} delves into analyzing the most probable configurations of these models. Section \ref{sec3} focuses on the study of fermion localization. In Section \ref{sec4}, we calculate probabilistic measures such as information entropy and relative probability. Finally, Section \ref{sec5} summarizes our key findings and presents concluding remarks.

\section{Braneworld-$f(Q)$}
\label{sec1}
In the framework of symmetric teleparallel gravity, we engage with a metric-affine geometry, delineating the metric and the affine connection as distinct entities. While the metric governs the notions of distances and angles within the space, the affine connection dictates covariant derivatives and the concept of parallel transport. Consequently, the general form of the affine connection is characterized by
\begin{equation}
\widetilde{\Gamma}^P\ _{MN}=\Gamma^P\ _{MN}+K^P\ _{MN}+L^P\ _{MN},   \end{equation}
where
\begin{equation}
\Gamma^P\ _{MN}=\frac{1}{2}g^{PS}\Big(\partial_M g_{SN}+\partial_N g_{SM}-\partial_S g_{MN} \Big)    
\end{equation}
is the Levi-Civita connection, and $K^{P}\ _{MN}$ is the contortion tensor which is described in terms of the torsion tensor $T^P\ _{MN}=\widetilde{\Gamma}^P\ _{NM}-\widetilde{\Gamma}^P\ _{MN}$, 
as
\begin{eqnarray}
K^{P}\ _{MN}=\frac{1}{2}\Big(T_{M}\ ^{P}\ _{N}+T_{M}\ ^{P}\ _{N}-T^{P}\ _{MN}\Big).
\end{eqnarray}
$L^P\ _{MN}$ is the distortion tensor which is defined as \cite{Nester:1998mp}
\begin{equation}
L^P\ _{MN}=\frac{1}{2}g^{PQ}\Big(Q_{PMN}-Q_{MPN}-Q_{NPM}\Big),   
\end{equation}
where
\begin{equation}
Q_{PMN}=\nabla_P g_{MN},   
\end{equation}
is the non-metricity tensor, which has two independent traces: $Q_M=g^{NP}Q_{MNP}$ and $\widetilde{Q}_M=g^{NP}Q_{NMP}$. 

Finally, we can define the dual non-metricity tensor (or conjugate non-metricity tensor) \cite{Nester:1998mp}
\begin{equation}
P^P\ _{MN}=-\frac{1}{2}L^P\ _{MN}+\frac{1}{4}(Q^P-\widetilde{Q}^P)g_{MN}-\frac{1}{8}(\delta^P_M Q_N+\delta^P_N Q_M),   
\end{equation}
which leads us to scalar non-metricity $Q=Q_{PMN}P^{PMN}$. 

It is noteworthy that the Lagrangian
\begin{equation}
\mathcal{L}=\frac{1}{2}\sqrt{-g}Q,
\end{equation}
generates the same equations of motion as general relativity. This is evident since the curvature is defined as
\begin{equation}
R=Q+\nabla_M(Q^M-\widetilde{Q}^M),    
\end{equation}
where $\nabla_M(Q^M-\widetilde{Q}^M)$ is nothing but a boundary term. Therefore, this model is also known as the STEGR.

A more general gravitational model is gravity $f(Q)$, which is an extension of STEGR. For a five-dimensional space , gravitational action is \cite{BeltranJimenez:2017tkd,BeltranJimenez:2019esp,BeltranJimenez:2019tme}
\begin{equation}\label{a1}
S=\int d^5x \sqrt{-g}\Big[\frac{1}{2k_g}f(Q)+\mathcal{L}_m\Big],
\end{equation} 
where $k_g=8\pi G$ is the gravitational constant and $\mathcal{L}_m$ is the matter Lagrangian. 

When we vary the action (\ref{a1}) in relation to the metric and the connection, we obtain the equation of motion
\begin{eqnarray}
\frac{2}{\sqrt{-g}}\nabla_K(\sqrt{-g}f_QP^K\ _{MN}) -\frac{1}{2}g_{MN}f+f_Q(P_{MKL}Q_N\ ^{KL}-2Q^ L\ _{KM}P^K\ _{NL})=k_g\mathcal{T}_{MN},\nonumber\\[2mm]
\nabla_M\nabla_N(\sqrt{-g}f_Q P_K\ ^{MN})=0,~~~~~
\end{eqnarray}
where $\mathcal{T}_{MN}$ is the momentum-energy tensor defined as
\begin{equation}
\mathcal{T}_{MN}=-2\frac{\delta \mathcal{L}_m}{\delta g^{MN}}+ g_{MN}\mathcal{L}_m.
\end{equation}
Here, we do $f\equiv f(Q)$, $f_Q\equiv \partial f/\partial Q$ for simplicity.

For a braneworld scenario, we will use the metric defined as
\begin{equation}\label{metric}
ds^2= e^{2A}\eta_{\mu\nu}dx^{\mu}dx^{\nu}+dy^2.    
\end{equation}
In this context, $\eta_{\mu\nu}$ denotes the Minkowski metric, while $e^{2A}$ stands for the warp factor controlling the brane's thickness, and $y$ symbolizes the extra dimension. This model delineates the brane as our familiar $4D$ world, existing within a larger space of extra dimensions ($5D$) known as the bulk. In this notation, uppercase Latin indices ($M,N,...=0,1,2,3,4$) denote the bulk coordinates, while Greek indices ($\mu,\nu,...=0,1,2,3$) represent the brane's coordinates.

Now, let's define the matter Lagrangian simply. Here, we'll focus on a single scalar field, denoted as the source $\phi(y)$, which solely depends on the extra dimension. This scalar field serves the purpose of determining the thickness of the brane.
\begin{equation}\label{scalar}
\mathcal{L}_m=-\frac{1}{2}\partial_M\phi\partial^M\phi-V(\phi).  
\end{equation}

Thus, the gravitational equations for the model are defined as
\begin{eqnarray}
\phi^{\prime\prime}+4A^{\prime}\phi^{\prime}&=&\frac{dV}{d\phi},\\
\label{erer}
12f_Q A^{\prime 2}-\frac{f}{2}&=&k_g\Big(\frac{\phi^{\prime 2}}{2}-V\Big),\\
\label{erer2}
-3\Big[f_Q (A^{\prime\prime}+4 A^{\prime 2})+ f_Q^{\prime}A^{\prime }\Big]&=&k_g\Big(\frac{\phi^{\prime 2}}{2}+V\Big).
\end{eqnarray}
Here, the prime ($^{\prime}$) denotes derivative with respect to extra dimension ($y$).

In pursuing our investigation of generalized Scalar-Tensor-Vector Gravity (STVG) models, we introduce two distinct gravity models denoted as $f(Q)$. The first, $f_1(Q) = Q + kQ^n$, and the second, $f_2(Q) = Q + k_1Q^2 + k_2Q^3$, are defined, where the parameters $n$ and $k$ serve to modify the conventional STVG gravitational framework. The selection of these models is grounded in their notable efficacy, as evidenced in various studies. These include analyses of the physical properties of a quintessence anisotropic stellar model \cite{Bhar:2023zwi}, investigations into the nature of dark energy \cite{Bhar:2023xku,Mussatayeva:2023aoa}, research in neutrino physics \cite{Atayde:2023aoj}, and the development of cosmological models \cite{Koussour:2023rly,Bajardi:2023vcc}. Furthermore, our approach will incorporate the use of specific ansatzes in our methodologies
\begin{align}\label{coreA}
A(y)&=-p\ln{\cosh(\lambda y)},
\end{align}
which is widely used in the study of thick branes \cite{Gremm1999,Moreira:2023pes,Bachas:2022etu,Zhao:2022ftm,Li:2022kly,Zhong:2022wlw}. Here the parameters $\lambda$ and $p$ control the width of the brane.

\subsection{Kink-like solutions}

For $f_1(Q)$ the equations (\ref{erer}) and (\ref{erer2}) become
\begin{eqnarray}
\label{22}\phi'^2(y)&=&-\frac{3}{k_g}\Big[1+12^{n-1}kn(2n-1)A'^{2(n-1)}\Big]A'',\\
\label{23}V(\phi(y))&=&-\frac{3}{2k_g}\Big[4A'^2+A''+12^{n-1}kn(2n-1)A'^{2(n-1)}(4A'^2+nA'')\Big].
\end{eqnarray}
The scalar field's configuration can be determined by solving Eq.(\ref{22}). It is crucial to emphasize that, for our model to hold physical significance, the scalar field's behavior must tend toward a constant value, $\phi_c$, asymptotically. Additionally, the potential (\ref{23}) needs to adhere to the condition $\partial V(\phi\rightarrow\pm\phi_c)/\partial \phi=0$. Notably, our model satisfies all these prerequisites.

\begin{figure}[ht!]
\begin{center}
\begin{tabular}{ccc}
\includegraphics[height=5cm]{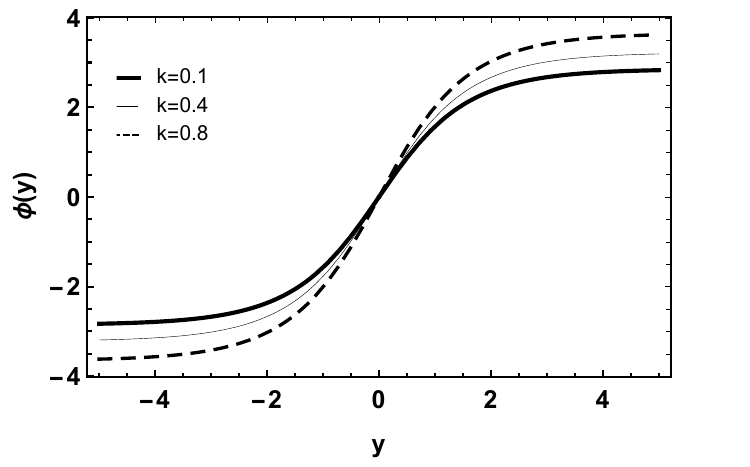}
\end{tabular}
\end{center}
\caption{ Kink-like solutions for $f_1$ with $\kappa_g=n=p=\lambda=1$. 
\label{fig1}}
\end{figure}

The dynamics of the $\phi$ field are particularly noteworthy. Referring to Fig.\ref{fig1}, we examine the behavior of the scalar field for the case when $n=1$. It is observed that the scalar field exhibits a kink-like solution, indicative of a domain wall formation at the core of the brane. This topological feature signifies a phase transition occurring within the brane. Moreover, as the value of the parameter $k$ is increased, the solution of the field becomes more accentuated. This implies significant alterations in both pressure and energy density on the brane.

\begin{figure}[ht!]
\begin{center}
\begin{tabular}{ccc}
\includegraphics[height=3.6cm]{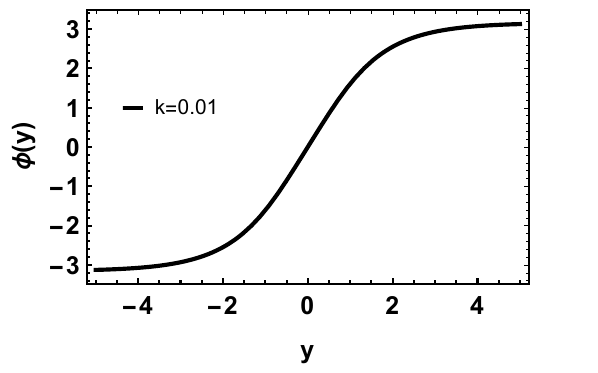}
\includegraphics[height=3.6cm]{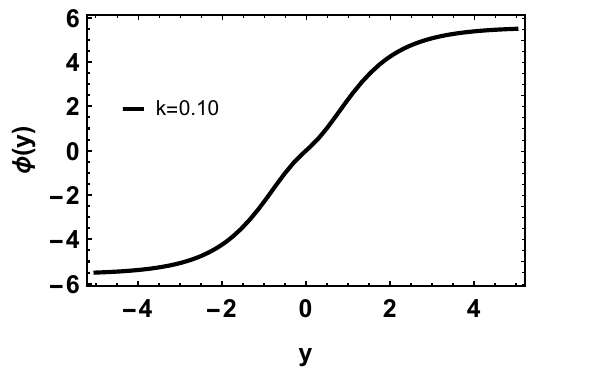}
\includegraphics[height=3.6cm]{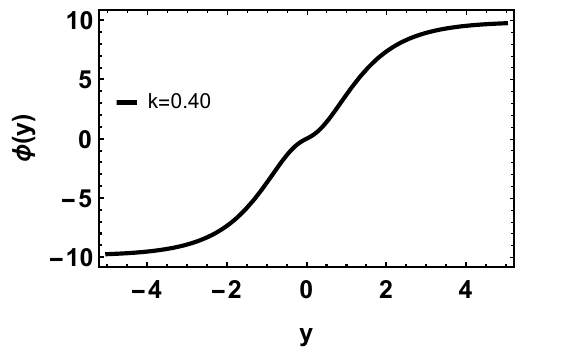}
\end{tabular}
\end{center}
\caption{ Kink-like solutions for $f_1$ with $n=2$ and $\kappa_g=p=\lambda=1$. 
\label{fig2}}
\end{figure}

In the case of $n=2$, an even more intriguing phenomenon occurs. As the value of $k$ is elevated, the previously observed kink-like solution transitions into a double-kink solution, as clearly illustrated in Fig.\ref{fig2}. This particular behavior symbolizes the emergence of additional phase transitions within the brane, effectively indicating the formation of new domain walls. Notably, such double-kink type solutions are relatively rare in the context of general relativity. The manifestation of this double-kink defect suggests a bifurcation within the brane structure, a phenomenon that should be reflected in the variations of both pressure and energy density on the brane.

\begin{figure}[ht!]
\begin{center}
\begin{tabular}{ccc}
\includegraphics[height=3.6cm]{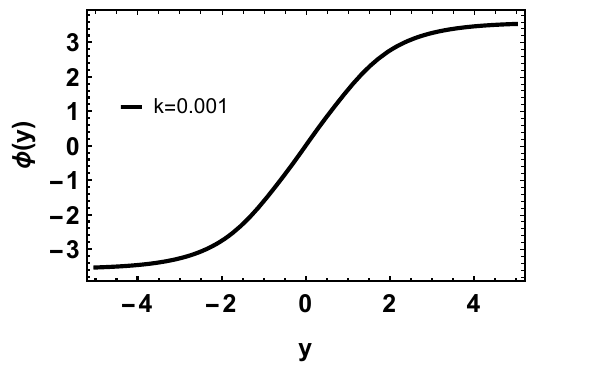}
\includegraphics[height=3.6cm]{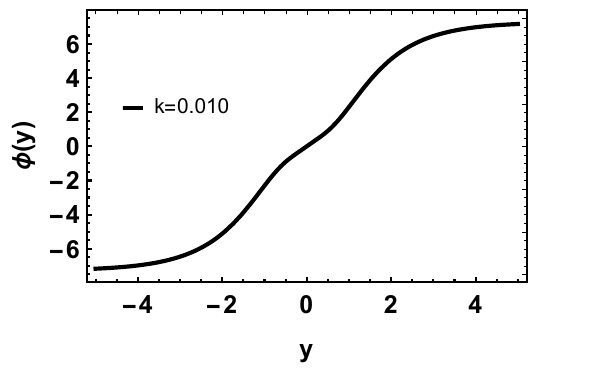}
\includegraphics[height=3.6cm]{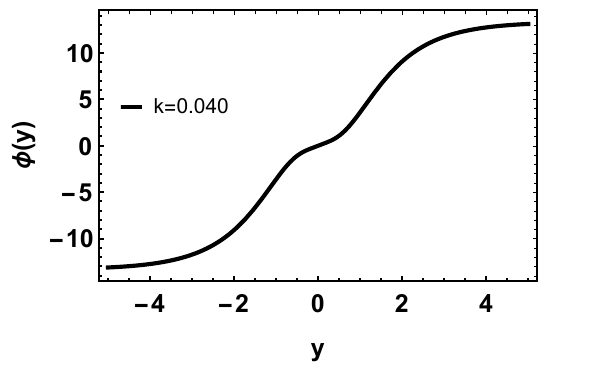}
\end{tabular}
\end{center}
\caption{ Kink-like solutions for $f_1$ with $n=3$ and $\kappa_g=p=\lambda=1$. 
\label{fig3}}
\end{figure}

Figure \ref{fig3} showcases the field solution's behavior for 
$n=3$. Once more, upon elevating the value of $k$, we observe the manifestation of a double-kink solution. However, in this scenario, the emergence of this structure is notably more pronounced compared to the $n=2$ case. Notably, the domain walls appear to be situated farther away from the core of the brane. This structural alteration may signify an uncommon density distribution of energy within the brane, suggesting the presence of three distinct density peaks, indicative of a more pronounced division within the brane structure.

For $f_2(Q)$, Eqs.(\ref{erer}) and (\ref{erer2}) become 
\begin{eqnarray}
\label{0.99}\phi'^2(y)&=&-\frac{3}{k_g}\Big[1+72A'^2(k_1+30k_2A'^{2})\Big]A'',\\
V(\phi(y))&=&-\frac{3}{2k_g}\Big\{A''4A'^2\Big[1+18k_1(2A'^2+A'')+180k_2(4A'^2+3A'²A'')\Big]\Big\}.
\end{eqnarray}

The physical viability of our model is contingent upon certain conditions being met, specifically, $\phi(y\rightarrow\pm\infty)=\pm\phi_c$ and $\partial V(\phi\rightarrow\pm\phi_c)/\partial \phi=0$. These conditions ensure that as the parameter $y$ approaches infinity, the scalar field $\phi$ reaches its limiting values of $\pm\phi_c$. Additionally, the derivative of the potential $V$ with respect to $\phi$ tends toward zero as $\phi$ approaches the limiting values $\pm\phi_c$. These conditions are essential to maintain the physical integrity and coherence of our model.

\begin{figure}[ht!]
\begin{center}
\begin{tabular}{ccc}
\includegraphics[height=3.6cm]{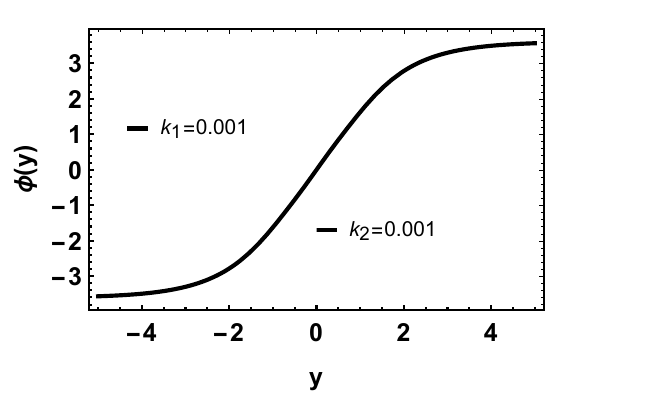}
\includegraphics[height=3.6cm]{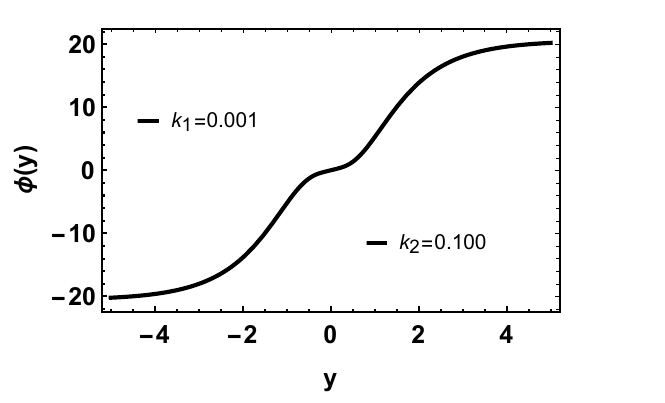}
\includegraphics[height=3.6cm]{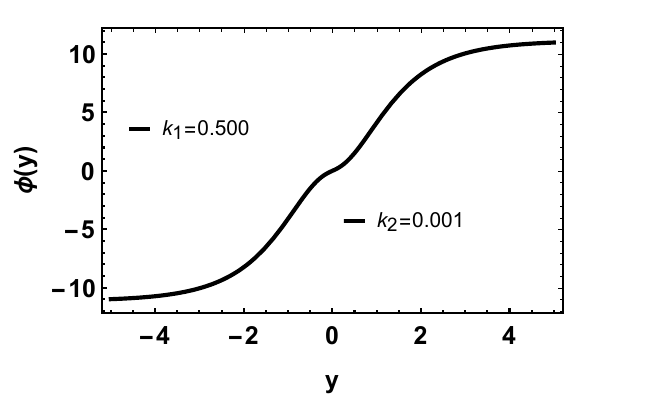}
\end{tabular}
\end{center}
\caption{ Kink-like solutions for $f_2$ with $\kappa_g=p=\lambda=1$. 
\label{fig4}}
\end{figure}

The solution for the field $\phi$ is obtained by solving Eq.(\ref{0.99}). Analyzing the behavior of the scalar field is made possible through Fig.\ref{fig4}. It's notable that the scalar field solution initially appears as a kink-like pattern and tends towards a double-kink configuration as we increase the values of $k_{1,2}$. This transition signifies the emergence of novel phase transitions, presenting new domain walls, and suggesting an inclination for the brane to split. This underscores how alterations in geometry, deviating from STEGR, can significantly impact the appearance of fresh structures leading to brane division. The manifestation of this splitting should become apparent when scrutinizing the energy densities within the brane.

\subsection{Energy density conditions}

Energy density is a pivotal parameter for understanding the behavior of the brane. Thus, in this subsection, our focus shifts towards examining the energy density residing on the brane, defined as:
\begin{align}\label{3333}
\rho(y)=-e^{2A}\mathcal{L}_m.
\end{align}
Through Eq.(\ref{erer2}) we can easily arrive at the form
\begin{eqnarray}
\rho(y)=-\frac{3}{k_g}\Big[2A'^2+A''+12^{n-1}k(2n-1)A'^{2(n-1)}(2A'^2+nA'')\Big]e^{-2A},
\end{eqnarray}
which is the energy density for $f_1(Q)$.

\begin{figure}[ht!]
\begin{center}
\begin{tabular}{ccc}
\includegraphics[height=5cm]{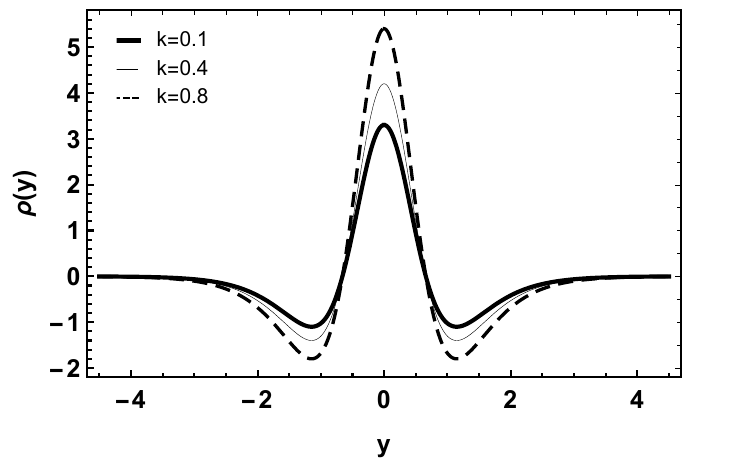}\\ 
(a) \\
\includegraphics[height=5cm]{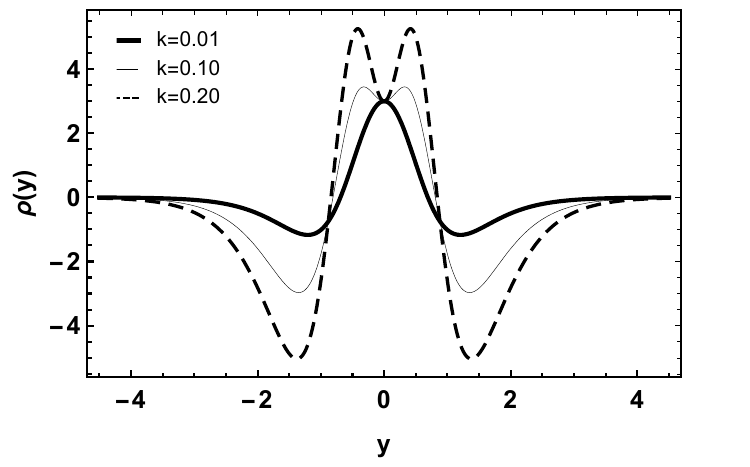}
\includegraphics[height=5cm]{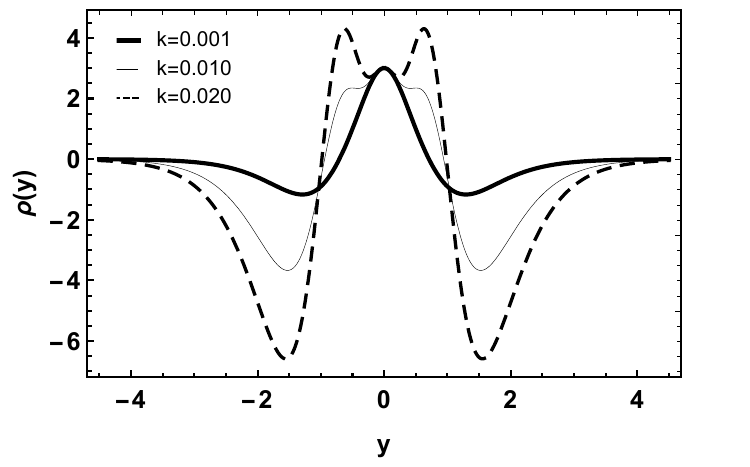}\\
(b) \hspace{6 cm}(c)
\end{tabular}
\end{center}
\caption{ Energy density for $f_1$ with $\kappa_g=p=\lambda=1$. (a) $n=1$. (b) $n=2$. (c) $n=3$.
\label{fig5}}
\end{figure}

In Fig.\ref{fig5}, we present the energy density behavior concerning 
$f_1(Q)$. For $n=1$ (Fig.\ref{fig5}.a), the energy density exhibits a distinct peak at the origin, signifying the precise localization of the brane within the system. As we manipulate the value of 
$k$, the amplitude of this peak intensifies. This outcome aligns consistently with the observations made in the preceding subsection. It is evident that the background scalar field's structure significantly influences the brane's behavior.

For the case of $n=2$, an intriguing phenomenon is observed in the energy density profile. As depicted in Fig.\ref{fig5}.b, when we increase the value of $k$, new density peaks begin to materialize near the origin. Additionally, the energy wells become more pronounced. This distinctive pattern is a direct consequence of the emergence of multiple phase transitions in the solution of the background scalar field. Due to the proximity of these phase transitions to the brane's core, only two new energy peaks are discernible. This scenario suggests a potential bifurcation of the brane.

In the $n=3$ scenario (Fig.\ref{fig5}.c), an even more unusual behavior is exhibited. The energy density shows a triad of density peaks as the value of $k$ is elevated. This behavior is indicative of the scalar field solution undergoing new phase transitions. The double-kink solution hints at possible segmentations within the brane, underscoring the impact of STEGR deviations on the emergence of novel structures within the brane's framework.

For $f_2(Q)$, the energy density has the form
\begin{eqnarray}
\rho(y)=-\frac{3}{k_g}\Big\{A''+2A'^2\Big[1+36k_1(A'^{2}+A'')+360k_2(2A'^4+3A'A'')\Big]\Big\}e^{-2A}.
\end{eqnarray}

\begin{figure}[ht!]
\begin{center}
\begin{tabular}{ccc}
\includegraphics[height=5cm]{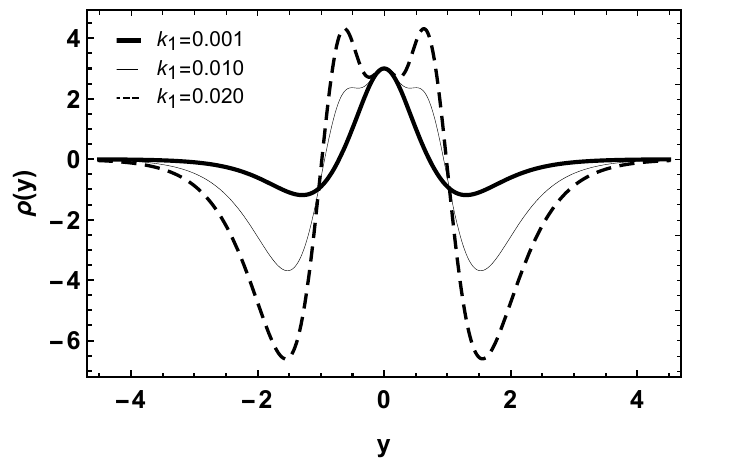}
\includegraphics[height=5cm]{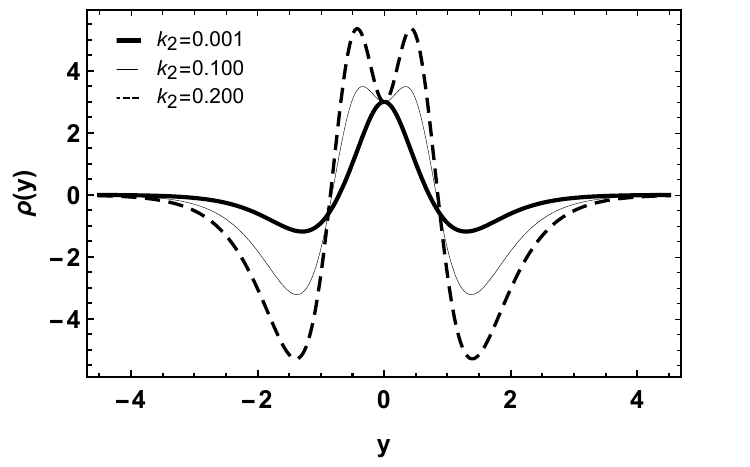}\\
(a) \hspace{6 cm}(b)
\end{tabular}
\end{center}
\caption{ Energy density for $f_2$ with $\kappa_g=p=\lambda=1$. (a) $k_2=0.001$. (b) $k_1=0.001$.
\label{fig6}}
\end{figure}

The energy density patterns for $f_{2}(Q)$ are illustrated in Fig.\ref{fig6}. As we manipulate the $k_1$ parameter, the emergence of new energy peaks becomes apparent. Specifically, two more pronounced peaks materialize around the core, accompanied by a less prominent peak at the origin. Upon further increasing the $k_2$ 
parameter, the single peak initially located at the origin splits into two distinct peaks surrounding this point. Moreover, the energy wells become more pronounced with escalating values of the $k_{1,2}$
parameters. These structural changes observed in the brane's energy density mirror the behaviors previously analyzed in the background scalar field subsection. This relationship underscores how the emergence of new phase transitions within the solution of the scalar field corresponds to the appearance of fresh peaks in energy density. Remarkably, the STEGR modification induces brane splitting dynamics employing only one background scalar field.

\section{Measure of probability and stability on the brane}
\label{sec2}

In this section, we delve into a succinct overview of DCE, an offshoot of configurational entropy \cite{GS}. Our focus will be to employ DCE as a tool for examining the stability of our model. In the context of braneworld scenarios, DCE has garnered considerable interest for its proficiency in identifying the most probable configuration of a given model and in predicting phase transitions, such as the formation of domain walls \cite{Correa2015a,Correa2015c,Correa2015b,Correa2016b,Correa2016a,Moreira:2021cta,Moreira:2022zmx}. This capability makes it an invaluable asset for analyzing complex physical systems where stability and phase behavior are key concerns.

In order to establish the Discrete Cosine Transform (DCT), it is imperative to utilize the Fourier transform of the energy density
\begin{align}
\mathcal{F}[\omega]=\frac{1}{\sqrt{2\pi}}\int e^{i\omega y}\rho(y) dy.
\end{align}
Recalling that the energy density is given by $\rho(y) = -e^{2A}\mathcal{L}_m$ (refer to Eq. (\ref{3333})), this leads us to the next step in our derivation
\begin{align}\label{4444}
\mathcal{F}[\omega]=-\frac{1}{\sqrt{2\pi}}\int e^{2A(y)+i\omega y}\mathcal{L}_m dy.
\end{align}

Using Eq.(\ref{4444}) we can define the modal fraction
\begin{align}
f(\omega)=\frac{\mid\mathcal{F}[\omega]\mid^2}{\int\mid\mathcal{F}[\omega]\mid^2d\omega}.
\end{align}
This expression represents the relative weight of each $\omega$ mode, always holding a value less than or equal to 1. Utilizing the modal fraction, we can then define the DCT as outlined in references \cite{GS,G1,G2,G3}
\begin{align}\label{dce_g}
S_C[f]=-\int \bar{f}(\omega)\ln[\bar{f}(\omega)]d\omega,
\end{align}
where $\bar{f}(\omega)=f(\omega)/f_{\text{max}}(\omega)$ signifies the normalized modal fraction, where $f_{\text{max}}(\omega)$ denotes the maximum value of this fraction. It is essential to note that the definition (\ref{dce_g}) is exclusively applicable to continuous functions within an open interval.

For the specific case of the gravity model $f_1(Q)$, when setting $n=1$ and considering the parameters $\kappa_g$, $p$, and $\lambda$ all equal to 1, the modal fraction takes on a distinct form
\begin{align}
f(\omega)=\frac{21}{128}\pi\omega^2\mathrm{csch}^2\Big(\frac{\pi\omega}{2}\Big).
\end{align}
It's worth noting that in this scenario, the modal fraction remains independent of the parameter $k$. Consequently, in this specific case, deriving the DCT concerning the non-metricity parameter becomes unattainable. However, when $n=2$, the modal fraction assumes the following form:
\begin{align}
f(\omega)=\frac{231\pi\omega^2[5\omega^2+2k(4-10\omega^2+\omega^4)]^2}{640[55+72k(114k+11)]}\mathrm{csch}^2\Big(\frac{\pi\omega}{2}\Big).
\end{align}
Indeed, when considering the case where $n=2$, the DCE exhibits dependence on the parameter $k$. Similarly, for $n=3$, this dependency is also observed, and the modal fraction is characterized by the following expression:
\begin{align}
f(\omega)=\frac{429\pi\omega^2[7\omega^2-2k(576-1400\omega^2+224\omega^4-5\omega^6)]^2}{896[143+96k(37800k+169)]}\mathrm{csch}^2\Big(\frac{\pi\omega}{2}\Big).
\end{align}

\begin{figure}[ht!]
\begin{center}
\begin{tabular}{ccc}
\includegraphics[height=5cm]{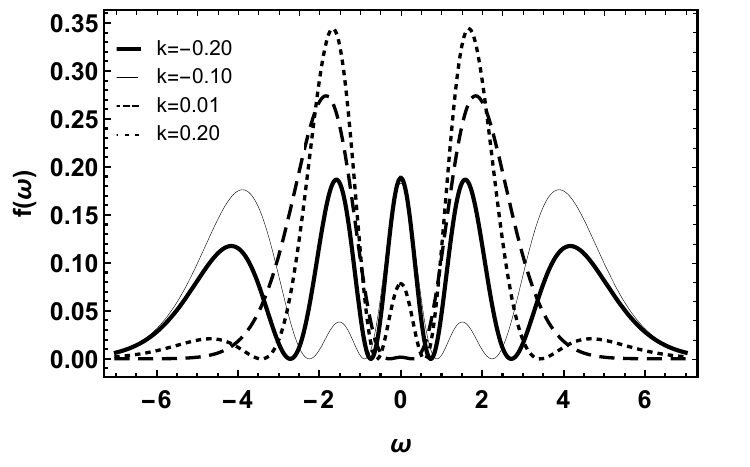}
\includegraphics[height=5cm]{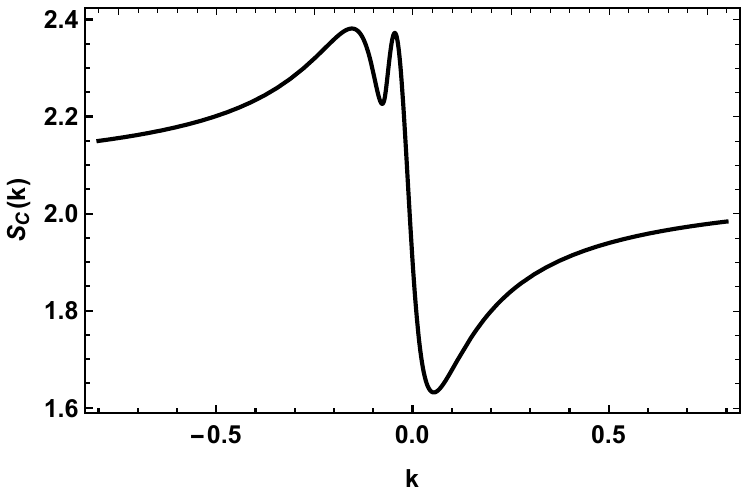}\\
(a) \hspace{6 cm}(b) \\
\includegraphics[height=5cm]{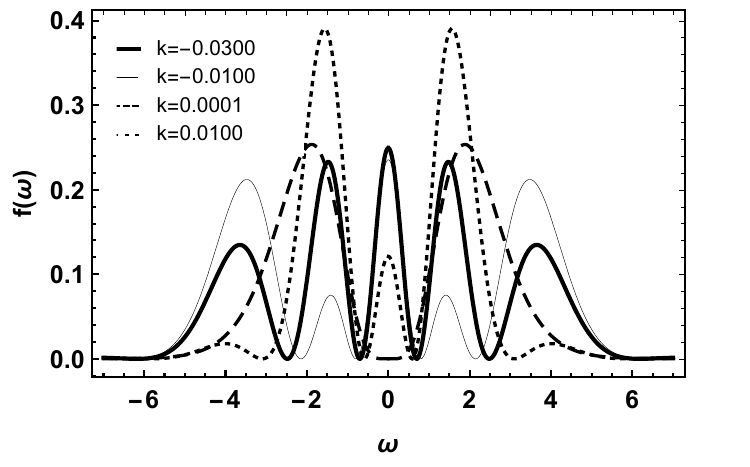}
\includegraphics[height=5cm]{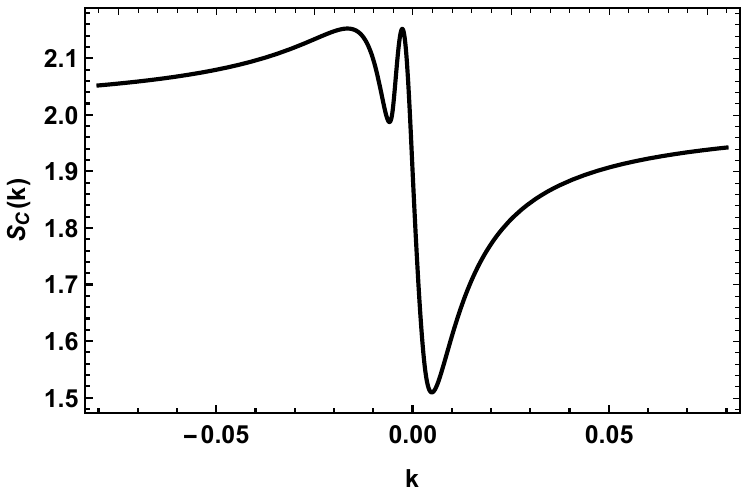}\\
(c) \hspace{6 cm}(d)
\end{tabular}
\end{center}
\caption{ For $f_1$ with $\kappa_g=p=\lambda=1$. Modal fraction (a) $n=2$ and (c) $n=3$. DCE (b) $n=2$ and (d) $n=3$.
\label{fig7}}
\end{figure}

The analysis of modal fraction and DCE for the $f_1(Q)$ model is depicted in Fig.\ref{fig7}. Notably, for $n=2$, changes in the value of $k$ lead to the emergence of new peaks in the modal fraction (Fig.\ref{fig7}.a). This phenomenon is directly linked to the stability of the model, as revealed through the DCE analysis. In Fig.\ref{fig7}.b, the DCE graph exhibits four extremal points, including two minima. The most pronounced minimum, occurring around $k\approx0.05$, signifies the point of greatest stability and the most probable configuration for the model.

For $n=3$, variations in $k$ similarly result in new peaks in the modal fraction (Fig.\ref{fig7}.c). Once more, the DCE graph shows two minimum points, but here, the most significant point of stability is located at $k\approx0.004$. This critical point highlights the transition from a single-phase (kink) to a double-phase (double-kink) transition. The presence of absolute minimum points in the DCE graph is especially noteworthy as it marks the evolutionary trajectory of the model from a singular to a dual phase transition state, underscoring the intricate interplay between the model parameters and the physical phenomena they represent.

In the model $f_2(Q)$ with $\kappa_g=p=\lambda=1$ the modal fraction is
\begin{align}
f(\omega)&=\frac{429\pi\omega^2[35\omega^2-84k_1(4-10\omega^2+\omega^4)-10k_2(576-1400\omega^2+224\omega^4-5\omega^6)]^2}{4480[715+106704k_1^2+81120k_2+18144000k_2^2+72k_1(34800k_2+143)]}\nonumber\\
&\times\mathrm{csch}^2\Big(\frac{\pi\omega}{2}\Big).
\end{align}
In this case, the modal fraction depends on the parameters that deviate from the usual STEGR.

\begin{figure}[ht!]
\begin{center}
\begin{tabular}{ccc}
\includegraphics[height=5cm]{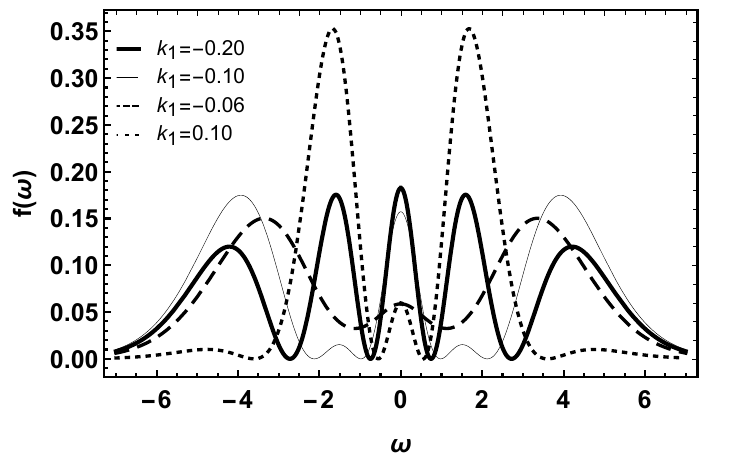}
\includegraphics[height=5cm]{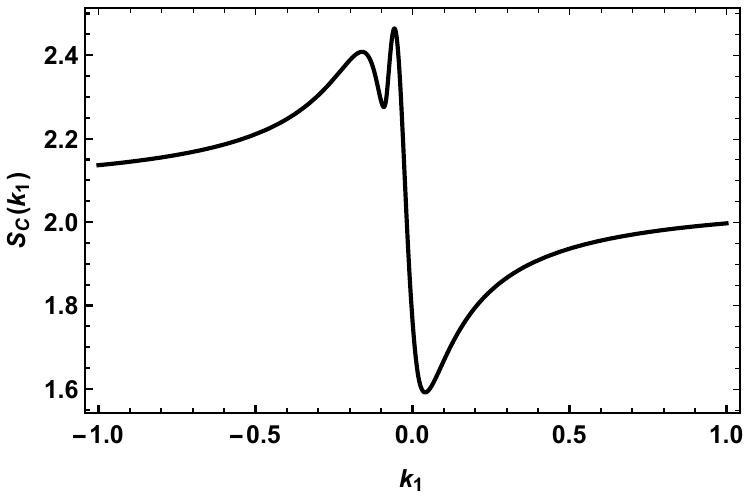}\\
(a) \hspace{6 cm}(b) \\
\includegraphics[height=5cm]{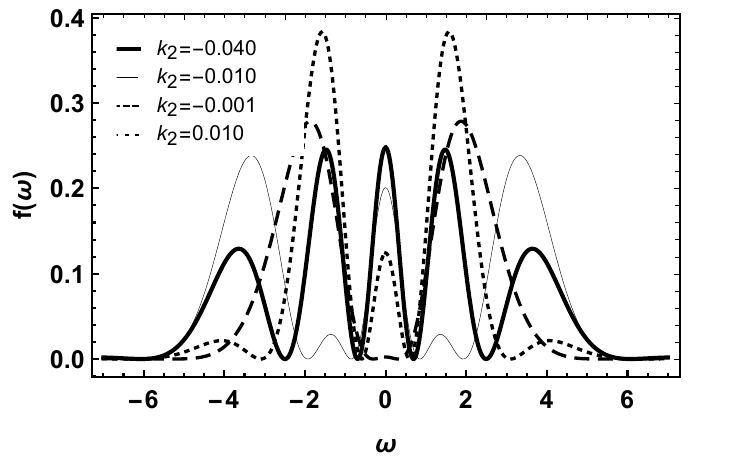}
\includegraphics[height=5cm]{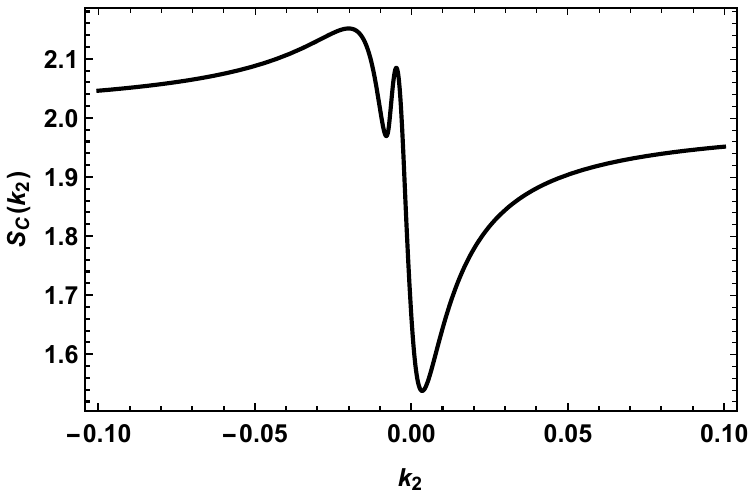}\\
(c) \hspace{6 cm}(d)
\end{tabular}
\end{center}
\caption{For $f_2$ with $\kappa_g=p=\lambda=1$. Modal fraction (a) $k_2=0.001$ and (c) $k_1=0.001$. DCE (b) $k_2=0.001$ and (d) $k_1=0.001$.
\label{fig8}}
\end{figure}

The analysis of the modal fraction and DCE for the $f_2(Q)$ model is presented in Fig.\ref{fig8}. Variations in the parameters $k_{1,2}$ result in the emergence of new peaks within the modal fraction (Fig.\ref{fig8}.a and c). Evaluating the stability of the model, the most pronounced extremal points in the DCE graph signify critical stability points. Notably, within the interval $0 < k_1 < 0.05$, a distinct minimum point stands out, representing the apex of stability for the system (Fig.\ref{fig8}.b). Similarly, around $k_2 \approx 0.002$ (Fig.\ref{fig8}.d), another sharp minimum point characterizes the most probable configuration of our model. These absolute minimum points delineate the evolution from a kink-like solution to a double-kink solution, signifying the emergence of novel structures within the brane. This correlation underscores how the model parameters influence the system's stability and the evolution of its structural configurations.

\section{Fermion localization}
\label{sec3}

In the preceding sections, we thoroughly explored how the parameters ($n$ and $k$), which govern deviations from standard STEGR gravity, influence the positioning of fields within the brane. To extend our understanding, this section delves into the localization of spin-1/2 fermions in the $f(Q)$ braneworld scenario.

For successful fermion localization on the brane, it is imperative to establish a coupling between the fermion field and the background scalar fields. This interaction is commonly referred to as Yukawa-type coupling, denoted by $\overline{\Psi}\phi\Psi$. Such a coupling mechanism has been extensively utilized in various studies due to its simplicity and effectiveness in precisely depicting fermion dynamics \cite{Liu2008,Liu2008b,Liu2009,Liu2009a,Liu2009b,Dantas2013,Moreira20211,Moreira:2023byr}.

In recent years, alternative coupling methodologies have emerged, such as the non-minimal coupling of fermions with the system's geometry \cite{Li:2017dkw,Guo:2019vvm,Moreira:2021wkj,Silva:2022pfd}. These novel approaches offer additional insights into the interaction mechanisms within the braneworld. However, the Yukawa-type coupling remains the most prevalent due to its relative simplicity and the robustness of the results it yields in modeling fermion behavior.

Thus, in our analysis, we will primarily focus on the Yukawa-type coupling to understand the localization dynamics of spin-1/2 fermions in the $f(Q)$ braneworld. This approach will facilitate a clearer understanding of how fermions behave under the influence of the background scalar fields and the modified gravitational dynamics characteristic of our braneworld model.

In the context of a minimal Yukawa coupling, the 5-dimensional Dirac action for particles with spin $1/2$ is given by the following expression:
\begin{eqnarray}\label{1}
\mathcal{S}_{1/2}=\int \sqrt{-g} \Big(\overline{\Psi}i\Gamma^M D_M\Psi -\xi \phi\overline{\Psi}\Psi\Big)d^5x.
\end{eqnarray}
where $\xi$ represents a dimensionless coupling constant. The term $D_M=\partial_M +\Omega_M$ denotes the covariant derivative, where $\Omega_M$ is the torsion-free spin connection. This particular spin connection is articulated in terms of the Levi-Civita connection, as defined in the following expression:
\begin{eqnarray}\label{3}
\Omega_M=\frac{1}{4}\Big(\Gamma_M\ ^{{\overline{N}}{\overline{Q}}}\Big)\ \Gamma_{\overline{N}}\Gamma_{\overline{Q}},
\end{eqnarray}
where $\Gamma^{{M}}$ are the curved Dirac matrices, which are defined through the plane Dirac matrices
($\Gamma^{\overline{M}}$), and \textit{vielbeins} ($E_{\overline{M}}\ ^M$), in the form
\begin{eqnarray}
\Gamma^M=E_{\overline{M}}\ ^M \Gamma^{\overline{M}}.
\end{eqnarray}
These matrices adhere to the Clifford algebra ${\Gamma^M,\Gamma^N}=2g^{MN}$. The "vielbeins" establish a tangent space and establish a connection to the metric via the expression $g_{MN}=\eta_{\overline{M}\overline{N}}E^{\overline{M}}_M E ^{\overline{N}}_N$, where the slashed capital Latin indices ($\overline{M},\overline{N},...=0,1,2,3,4$) represent the coordinates within the tangent space. For ease of representation, we undertake the transformation $dz=e^{-A(y)}dy$, leading the metric (\ref{metric}) to assume the form $ds^ 2=e^{2A}(\eta ^{\mu\nu}dx^\mu dx^\nu+dz^2)$.

The Dirac equation (\ref{1}) takes the form
\begin{eqnarray}\label{7}
\Big[\gamma^{\mu}\partial_\mu+\gamma^4(\partial_z+2\dot{A})-\xi e^A\phi\Big]\psi=0,
\end{eqnarray}
where
\begin{eqnarray}
\Psi\equiv\Psi(x,z)=\left(\begin{array}{cccccc}
\psi\\
0\\
\end{array}\right),\ 
\Gamma^{\overline{\mu}}=\left(\begin{array}{cccccc}
0&\gamma^{\overline{\mu}}\\
\gamma^{\overline{\mu}}&0\\
\end{array}\right),\ \Gamma^{\overline{z}}=\left(\begin{array}{cccccc}
0&\gamma^4\\
\gamma^4&0\\
\end{array}\right),
\end{eqnarray}
is the representation of the spinor and dot (\ $\dot{ }$\ ) denotes derivative with respect to extra dimension ($z$).

Performing the Kaluza-Klein decomposition on spinor
\begin{eqnarray}
\psi=\sum_n[\psi_{L,n}(x)\varphi_{L,n}(z)+\psi_{R,n}(x)\varphi_{R,n}(z)],
\end{eqnarray}
we arrive at the coupled equations
\begin{eqnarray}\label{9}
\Big[\partial_z+\xi e^A \phi\Big]\varphi_{L}(z)&=&m \varphi_{R}(z),\nonumber\\
\Big[\partial_z-\xi e^A\phi\Big]\varphi_{R}(z)&=&-m \varphi_{L}(z).
\end{eqnarray}
In this context, $\gamma^4\psi_{R,L}=\pm\psi_{R,L}$ represents the left-handed ($\psi_L$) and right-handed ($\psi_R$) components originating from the Dirac field. Additionally, the condition $\gamma^\mu\partial_\mu\psi_{R,L}=m\psi_{L,R}$ is satisfied.

We can easily decouple Eqs.(\ref{9}) and arrive at Schr\"{o}dinger-like equations
\begin{eqnarray}\label{10}
\Big[-\partial^2_z+V_L(z)\Big]\varphi_{L}(z)&=&m^2 \varphi_{L}(z),\nonumber\\
\Big[-\partial^2_z+V_R(z)\Big]\varphi_{R}(z)&=&m^2 \varphi_{R}(z).
\end{eqnarray}
In this framework, the effective potential is defined as $V_{R,L}(z) = U^2 \pm\partial_{z}U$, where $U = \xi e^A \phi$ represents the superpotential. It is important to note that Eq. (\ref{10}) adopts the form characteristic of supersymmetric quantum mechanics (SUSY-type). This formulation ensures the absence of tachyonic Kaluza-Klein (KK) states. Moreover, the inherent structure of supersymmetry facilitates the existence of a well-localized massless mode, particularly when $\xi > 0$.

\begin{eqnarray}
\varphi_{R0,L0}(z)\propto e^{\pm\int Udz}.
\end{eqnarray}

The localization of massive modes can be achieved through numerical solutions by imposing specific boundary conditions, as described in previous works \cite{Liu2009,Liu2009a,Moreira20211,Moreira:2021wkj}.
\begin{eqnarray}\label{ade}
\varphi_{even}(0)&=&c,\ \  \partial_z\varphi_{even}(0)=0,\nonumber\\ \varphi_{odd}(0)&=&0, \ \  \partial_z\varphi_{odd}(0)=c.
\end{eqnarray}
We use these boundary conditions since the effective potential $V_{R,L}(z)$ are even functions. Furthermore, the conditions (\ref{ade}) guarantee that the solutions $\varphi_{R,L}(z)$ will have the behavior of even $\varphi_{even}$ or odd wave functions $\varphi_{odd }$.

\begin{figure}[ht!]
\begin{center}
\begin{tabular}{ccc}
\includegraphics[height=5cm]{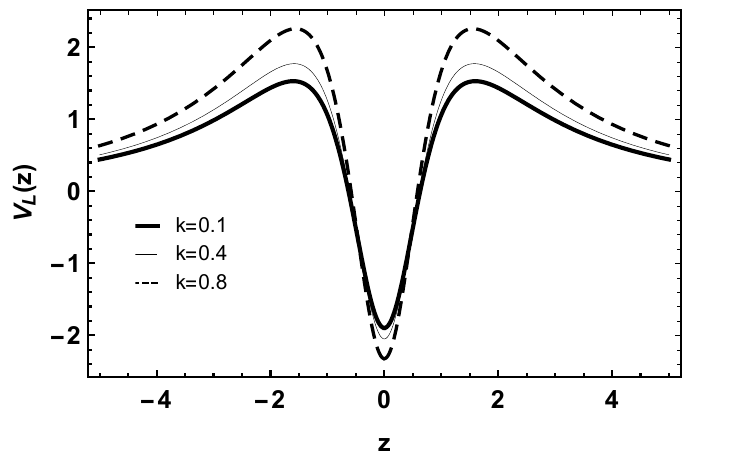}
\includegraphics[height=5cm]{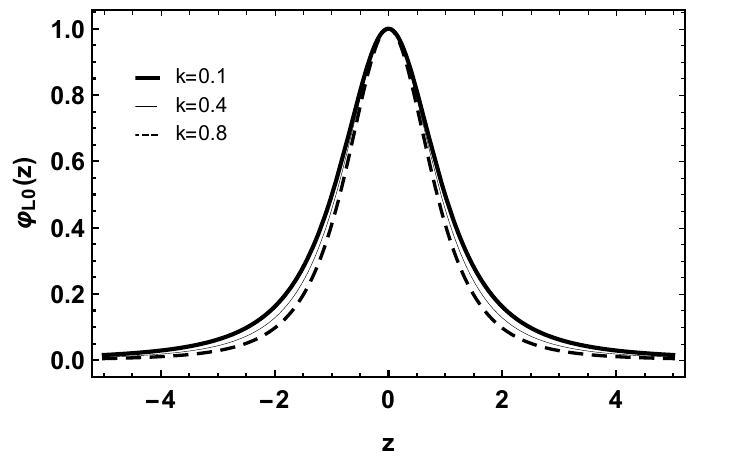}\\
(a) \hspace{6 cm}(b) \\
\includegraphics[height=5cm]{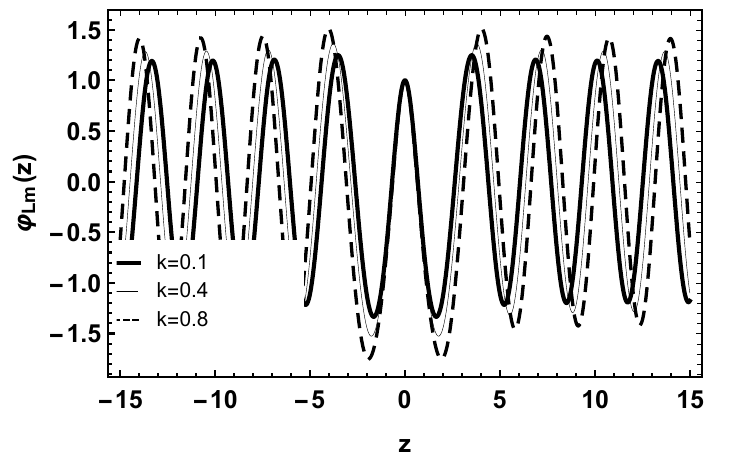}
\includegraphics[height=5cm]{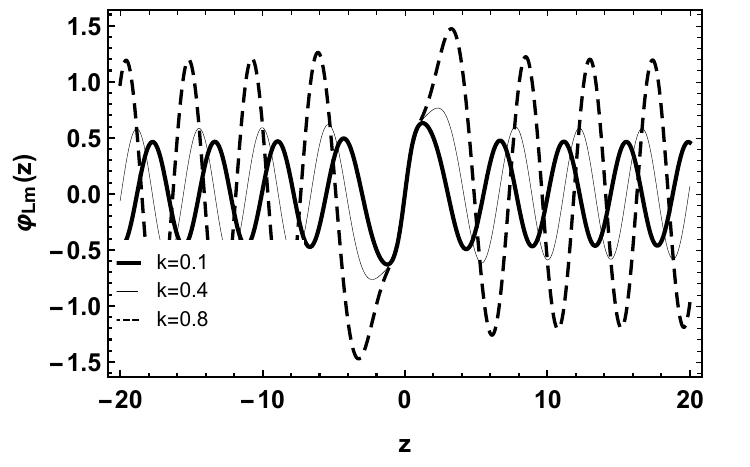}\\
(c) \hspace{6 cm}(d)
\end{tabular}
\end{center}
\caption{ For $f_1$ with $\kappa_g=n=p=\lambda=1$. (a) Effective potential. (b) Massless mode. Massive modes (c) even $m=2.063$ and (d) odd $m=1.532$.
\label{fig9}}
\end{figure}

In the $f_1(Q)$ model with $n=1$, the effective potential takes the form of a confining well at the origin, surrounded by two smaller barriers. With the variation of the parameter $k$, both the well and the barriers become more pronounced, as shown in Fig.\ref{fig9}.a. Massless modes are increasingly localized with the increase in $k$, as depicted in Fig.\ref{fig9}.b. For massive modes, the solutions resemble free waves with amplitudes that become steeper near the origin, intensifying as $k$ increases, as seen in Fig.\ref{fig9}.c and Fig.\ref{fig9}.d.

\begin{figure}[ht!]
\begin{center}
\begin{tabular}{ccc}
\includegraphics[height=5cm]{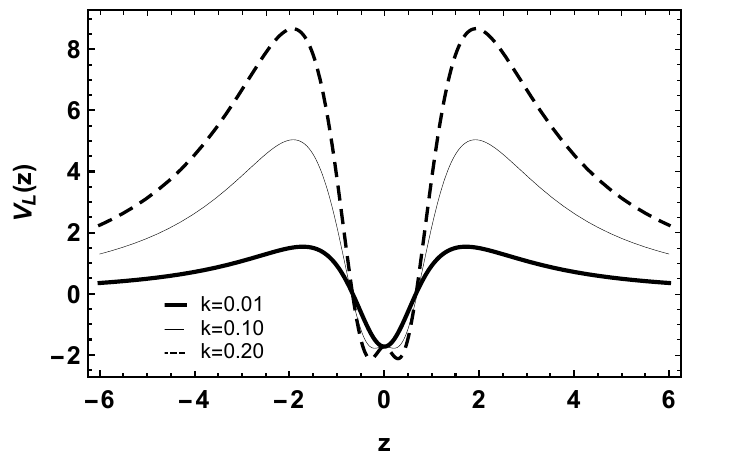}
\includegraphics[height=5cm]{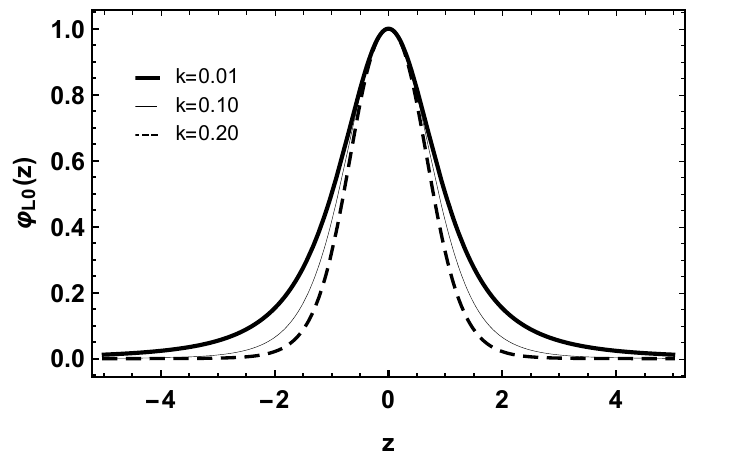}\\
(a) \hspace{6 cm}(b) \\
\includegraphics[height=5cm]{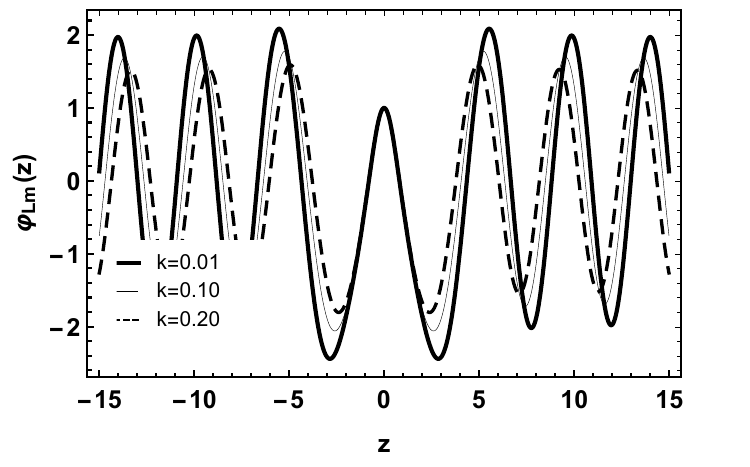}
\includegraphics[height=5cm]{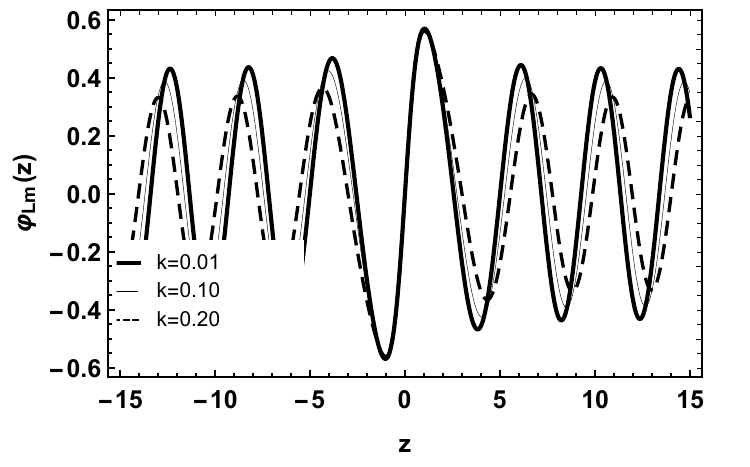}\\
(c) \hspace{6 cm}(d)
\end{tabular}
\end{center}
\caption{ For $f_1$ with $n=2$ and $\kappa_g=p=\lambda=1$. (a) Effective potential. (b) Massless mode. Massive modes (c) even $m=1.841$ and (d) odd $m=1.374$.
\label{fig10}}
\end{figure}

In the scenario where $n=2$, an interesting phenomenon occurs as the value of $k$ is increased: the effective potential well undergoes a division into two separate wells, as shown in Fig.\ref{fig10}.a. This division significantly impacts the behavior of massless modes, causing them to exhibit increased localization, as evident in Fig.\ref{fig10}.b. Additionally, this variation influences the behavior of massive modes, leading to oscillations with reduced amplitudes, depicted in Fig.\ref{fig10}.c and d. Notably, this division in the effective potential well occurs approximately around the value of $k \approx 0.05$, which coincides with the stability point observed in the preceding section.

\begin{figure}[ht!]
\begin{center}
\begin{tabular}{ccc}
\includegraphics[height=5cm]{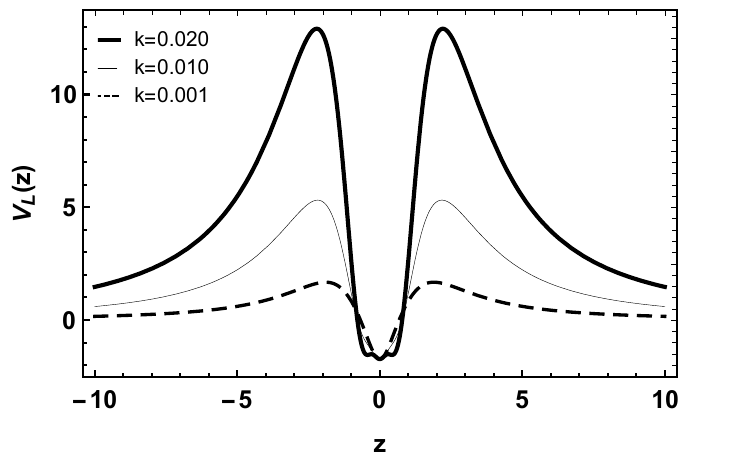}
\includegraphics[height=5cm]{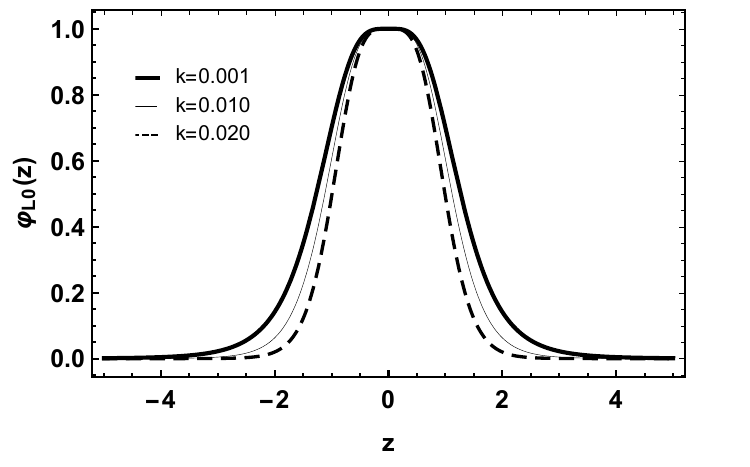}\\
(a) \hspace{6 cm}(b) \\
\includegraphics[height=5cm]{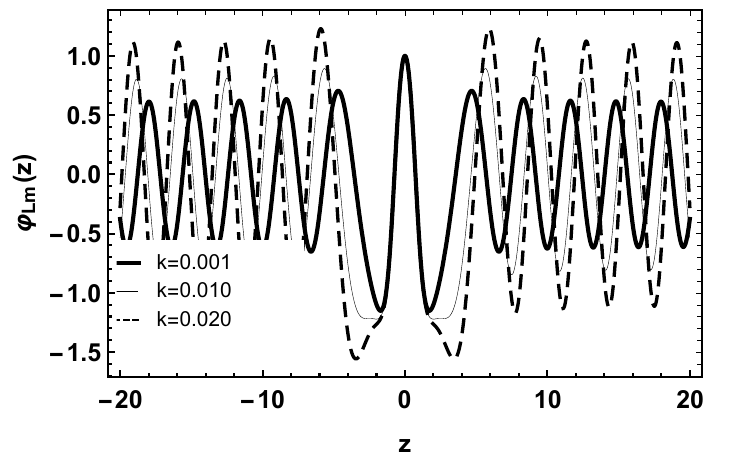}
\includegraphics[height=5cm]{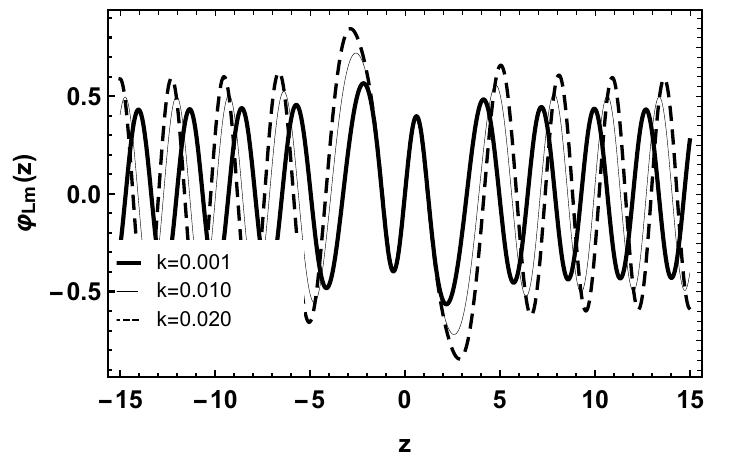}\\
(c) \hspace{6 cm}(d)
\end{tabular}
\end{center}
\caption{ For $f_1$ with $n=3$ and $\kappa_g=p=\lambda=1$. (a) Effective potential. (b) Massless mode. Massive modes (c) even $m=2.026$ and (d) odd $m=1.363$.
\label{fig11}}
\end{figure}

For $n=3$ in the $f_1(Q)$ model, increasing the value of $k$ reshapes the effective potential into a well with three distinct minima, as shown in Fig.\ref{fig11}.a, while also strengthening the surrounding potential barriers. This change leads to massless modes exhibiting a flattened peak (Fig.\ref{fig11}.b), which become more localized with higher $k$ values. The massive modes display interesting oscillation patterns near the brane core, with pronounced amplitudes that are less influenced by changes in $k$ (Fig.\ref{fig11}.c and d). Oscillations further from the core, however, do show increased amplitudes with rising $k$. Notably, the division of the potential well aligns with the model's stability point around $k\approx0.04$, indicating the most likely configuration of the system.

\begin{figure}[ht!]
\begin{center}
\begin{tabular}{ccc}
\includegraphics[height=5cm]{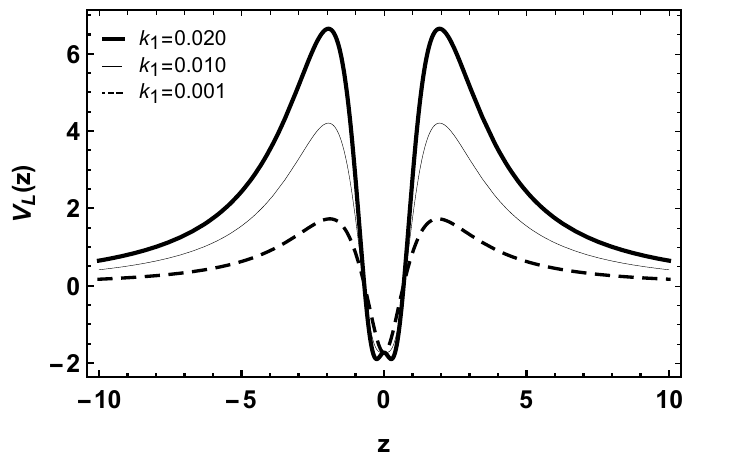}
\includegraphics[height=5cm]{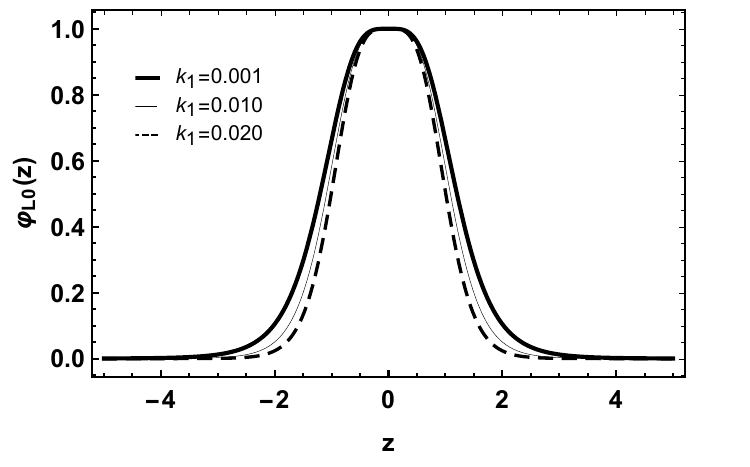}\\
(a) \hspace{6 cm}(b) \\
\includegraphics[height=5cm]{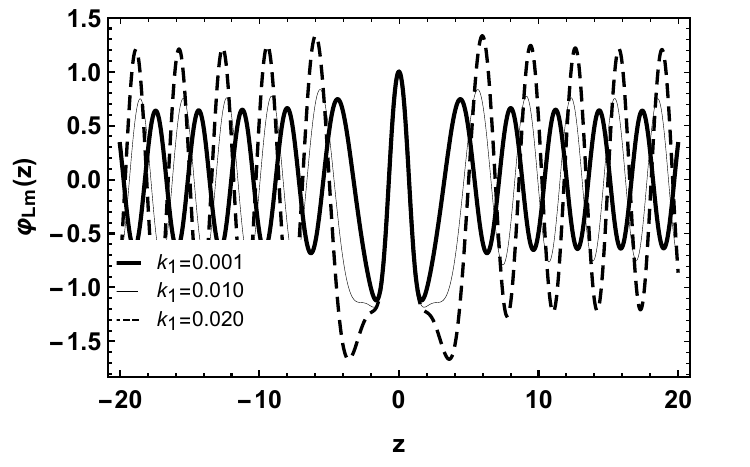}
\includegraphics[height=5cm]{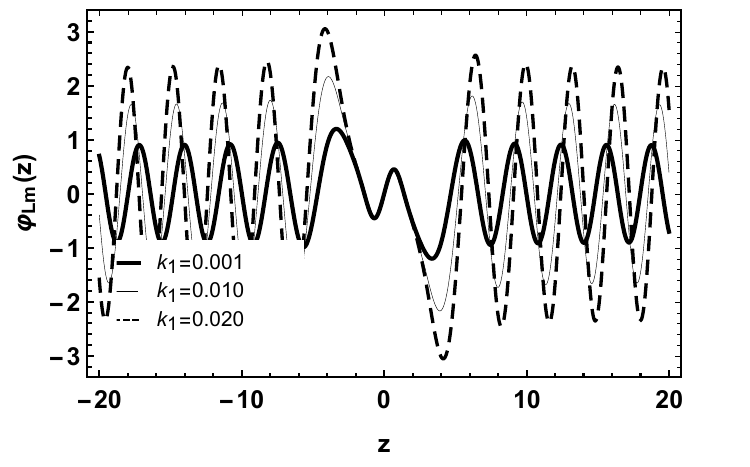}\\
(c) \hspace{6 cm}(d)
\end{tabular}
\end{center}
\caption{ For $f_2$ with $k_2=0.001$ and $\kappa_g=p=\lambda=1$. (a) Effective potential. (b) Massless mode. Massive modes (c) even $m=2.060$ and (d) odd $m=1.401$.
\label{fig12}}
\end{figure}

For the $f_2(Q)$ model, the effective potential presents a well that tends to split at the origin when we increase the value of $k_1$ (Fig.\ref{fig12}.a). Massless modes sense the change in effective potential, tending to become more localized. Furthermore, the massless mode has a flat peak, due to the shape of the potential (Fig.\ref{fig12}.b). The massive modes also present an interesting behavior, where the amplitudes in the brane core do not change with the variation of $k_1$. Only the oscillations that are not at the origin have their amplitudes intensified as $k_1$ increases (Fig.\ref{fig12}.c and d). Note that, in the interval $0<k_1<0.05$, the effective potential well is divided.

\begin{figure}[ht!]
\begin{center}
\begin{tabular}{ccc}
\includegraphics[height=5cm]{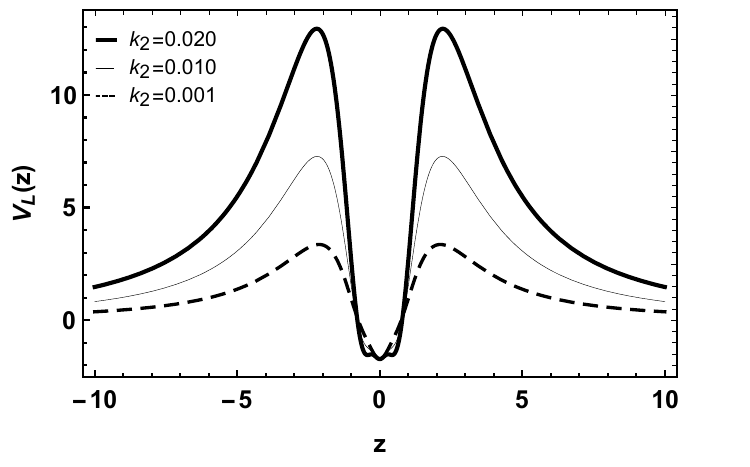}
\includegraphics[height=5cm]{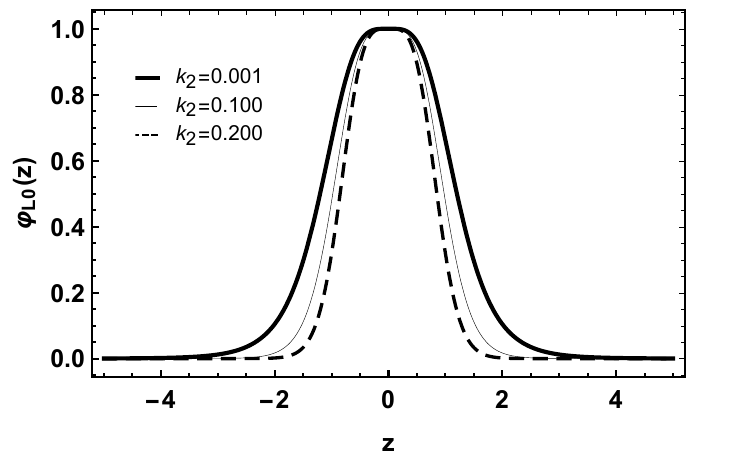}\\
(a) \hspace{6 cm}(b) \\
\includegraphics[height=5cm]{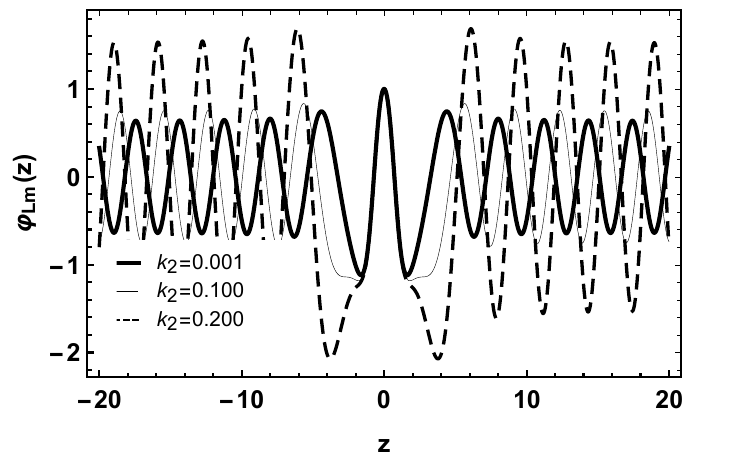}
\includegraphics[height=5cm]{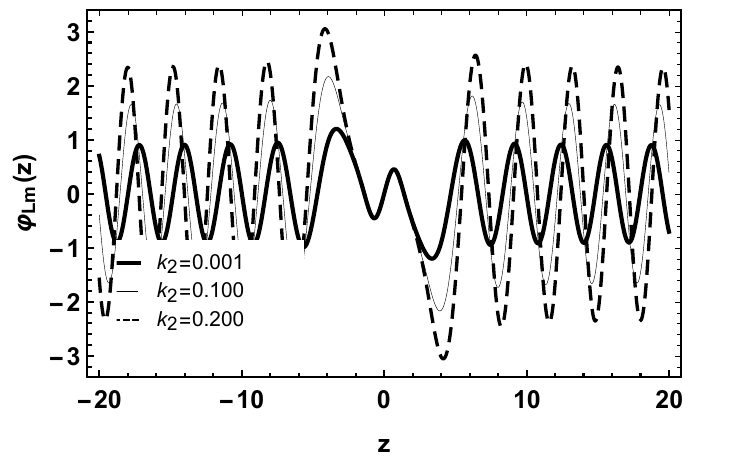}\\
(c) \hspace{6 cm}(d)
\end{tabular}
\end{center}
\caption{ For $f_2$ with $k_1=0.001$ and $\kappa_g=p=\lambda=1$. (a) Effective potential. (b) Massless mode. Massive modes (c) even $m=2.060$ and (d) odd $m=1.401$.
\label{fig13}}
\end{figure}

In the $f_2(Q)$ model, varying $k_2$ leads to a division of the effective potential well into three minima at the origin, alongside more pronounced potential barriers (Fig.\ref{fig13}.a). Massless modes become more localized with a flat peak in the brane's core (Fig.\ref{fig13}.b). Massive modes experience increased amplitude oscillations; however, the peak at the origin remains consistent despite variations in $k_2$ (Fig.\ref{fig13}.c and d). This division in the potential well correlates with the model's stability point near $k_2\approx0.002$ and signifies a split in the brane.

\section{Probability measures at localization}
\label{sec4}

To enhance our understanding of fermion localization on the brane, this section will focus on calculating two key probabilistic measures: the information entropy concerning massless fermionic modes and the relative probability associated with massive modes. These measures will provide a more detailed analysis of fermion localization conditions within the brane structure.

\subsection{Information entropy}

Quantum information theories have significantly influenced various physical systems, primarily due to their exceptional capability in offering a nuanced analysis of particle locations via the system's probability density. A fundamental development in information theory was Shannon's entropy theory \cite{Shannon}, which focuses on evaluating information loss or noise in the communication process between a sender and a receiver. Although initially a rudimentary concept, Shannon's theory has been instrumental in revolutionizing modern technology and continues to play a pivotal role in the advancement of quantum computing. Intriguingly, this measure of information can be adapted to examine the localization of massless fermionic modes in a braneworld context. In the following discussion, we will first introduce some basic concepts of Shannon entropy before applying these principles to our specific model.

To define Shannon entropy we need to apply the Fourier transform to the massless mode function, in the form
\begin{equation}\label{fouu}
\vert\varphi_{L0,R0}(p_z)\vert^{2}=\frac{1}{\sqrt{2\pi}}\int_{-\infty}^{\infty}\vert\varphi_{L0,R0}(z)\vert^{2}\,\text{e}^{-ipz} dz,
\end{equation}
where the variable $p_z$ represents the coordinate within the momentum space, also known as reciprocal space. Utilizing Eq. (\ref{fouu}), we can establish the Shannon entropy for both position space and momentum space in the following manner:
\begin{eqnarray}\label{0.11}
S_{z}&=&-\int_{-\infty}^{\infty}\vert\varphi_{L0,R0}(z)\vert^{2}\ln\vert\varphi_{L0,R0}(z)\vert^{2}dz,\nonumber\\
S_{p_z}&=&-\int_{-\infty}^{\infty}\vert\varphi_{L0,R0}(p_z)\vert^{2}\ln\vert\varphi_{L0,R0}(p_z)\vert^{2}dz.
\end{eqnarray}

The entropic measurements (\ref{0.11}) offer us an uncertainty relation known as the BBM relation \cite{Beckner,Bialy}, named after its proposers: Beckner, Bialynicki-Birula, and Mycielski. This entropic uncertainty relationship presents a compelling alternative to the Heisenberg uncertainty principle. The BBM uncertainty relation is defined as:
\begin{equation}
S_{z}+S_{p_z}\geq D(1+\text{ln}\pi).
\end{equation}

In this context, $D$ signifies the number of dimensions sensitive to changes in the system's information. Within our model, only the extra dimension is responsive to the entropic alterations of the system, hence $D=1$.

The numerical determination of Shannon entropy, as shown in Tables \ref{tab1} and \ref{tab2}, reveals key insights into the entropic information measurements for the two models $f_{1,2}$. In the $f_1$
model, an increase in the parameter $k$ correlates with a decrease in information entropy in position space, suggesting a higher probability of precisely locating the fermion and thus reduced uncertainties about its position. Conversely, in momentum space, information entropy increases with $k$, indicating a tendency for the fermion to be less localized and implying greater uncertainties in its momentum.

This trend becomes more pronounced with higher values of $n$. Additionally, the total information entropy of the system conforms to the BBM relation, underscoring that as $k$ and $n$ increase, so do the overall uncertainties in locating the fermion. This outcome is particularly significant as it implies that deviations from conventional STEGR are associated with increased uncertainties in pinpointing the fermion's position on the brane.

\begin{table}[h!]
\centering
\begin{tabular}{|c|c|c|c|c|c|}
\hline
$n$ & $k$ & $S_{z}$ & $S_{p_z}$ & $S_{z}+S_{p_z}$ & $1+\ln\pi$\\ \hline
\hline
  & 0.1  & 1.08361 & 1.08685  & 2.17046 & \\
1 & 0.4 & 0.98318 & 1.18938  & 2.17256 & 2.14473 \\
  & 0.8 & 0.88482 & 1.29251  & 2.17733 &\\ \hline \hline
  & 0.1  & 0.83056 & 1.40716  & 2.23772 & \\
2 & 0.2  & 0.72506 & 1.52045  & 2.24551 & 2.14473 \\
  & 0.4 & 0.61037 & 1.64310  & 2.25347 &\\ \hline \hline
  & 0.01  & 1.09926 & 1.18558  & 2.28484 & \\
3 & 0.02  & 0.94864 & 1.33664  & 2.28528 & 2.14473 \\
  & 0.04 & 0.81261 & 1.47385  & 2.28646 &\\ \hline  
\end{tabular}\\
\caption{Shannon's entropy for $f_1$ with $\kappa_g=\xi=p=\lambda=1$.\label{tab1}}
\end{table}

In the $f_2$  model, the trends observed in entropic information measurements exhibit a distinct relationship with the parameters $k_{1,2}$. Specifically, as the values of $k_{1,2}$ increase, there is a decrease in the measures of entropic information in position space. This suggests a heightened certainty in determining the fermion's location within the brane. In contrast, in momentum space, the information entropy displays an increasing trend with $k_{1,2}$. This implies that the fermion becomes less localized in momentum space as $k_{1,2}$ values rise, leading to greater uncertainties in its momentum.

Moreover, a notable increase in the total entropy is observed as $k_{1,2}$ values escalate, indicating a decrease in the overall certainty of the fermion's location. This behavior mirrors the patterns seen in the $f_1$  model and leads to an important conclusion: greater deviations from the standard gravitational model correlate with less certainty in pinpointing the fermion's location within the brane. This finding underscores a fundamental connection between the modifications in the gravitational model and the resulting uncertainties in fermion localization.

\begin{table}[h!]
\centering
\begin{tabular}{|c|c|c|c|c|c|}
\hline
$k_1$ & $k_2$ & $S_{z}$ & $S_{p_z}$ & $S_{z}+S_{p_z}$ & $1+\ln\pi$\\ \hline
\hline
0.01  & 0.01 & 1.09172 & 1.08961  & 2.18133 &  2.14473\\ \hline \hline
0.04  &      & 1.07073 & 1.11095  & 2.18168 &  \\
0.08  & 0.01 & 1.04591 & 1.13855  & 2.18446 &2.14473\\ 
0.20  &      & 0.98663 & 1.19799  & 2.18462 & \\ \hline \hline
      & 0.04 & 0.81106 & 1.37907  & 2.19013 &  \\
0.01  & 0.08 & 0.68637 & 1.51164  & 2.19801 &2.14473\\
      & 0.20 & 0.53260 & 1.69173  & 2.22433 & \\ \hline  
\end{tabular}\\
\caption{Shannon's entropy for $f_2$ with $\kappa_g=\xi=p=\lambda=1$.\label{tab2}}
\end{table}

\subsection{Relative probability}

Another frequently employed probability measure in determining fermion localization on branes is the relative probability, as documented in various references \cite{Liu2009,Liu2009a,Mendes:2017hmv,Guerrero:2019qqj,Xie:2019jkq,Guo:2019vvm}. What makes this measurement particularly intriguing is its application to non-localized functions, specifically massive fermionic modes, which possess a high probability of extending beyond the confines of the brane. The relative probability is defined as follows:
\begin{eqnarray}
P(m)=\frac{\int_{-z_b}^{z_b} \vert\varphi_{L,R}(z)\vert^2 dz}{\int_{-z_{max}}^{z_{max}} \vert\varphi_{L,R}(z)\vert^2 dz}.
\end{eqnarray}
In this context, $z_b$ represents a narrow band, while $z_{max}$ signifies the domain boundary. The relative probability measure serves to assess the likelihood of locating a particle with mass $m$ within a narrow band of $2z_b$. It helps to identify the massive fermionic modes that exhibit higher amplitudes near the brane core, known as resonant modes. These resonant modes denote the massive fermion modes most likely to localize on the brane. Let's proceed to identify these resonant modes within our model.

\begin{figure}[ht!]
\begin{center}
\begin{tabular}{ccc}
\includegraphics[height=5cm]{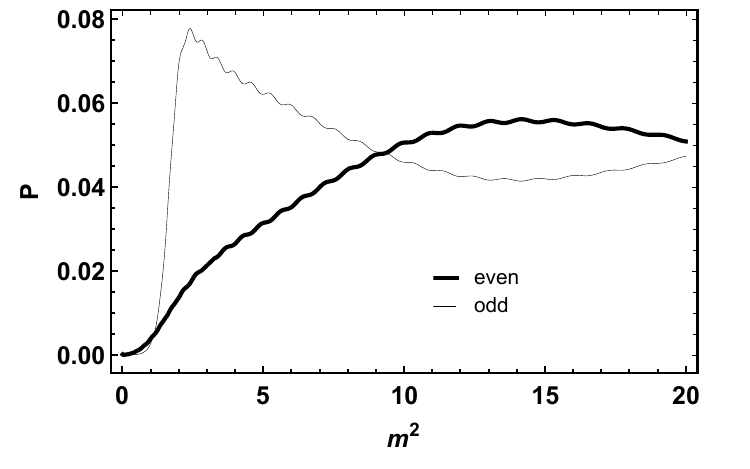}\\ 
(a) \\
\includegraphics[height=5cm]{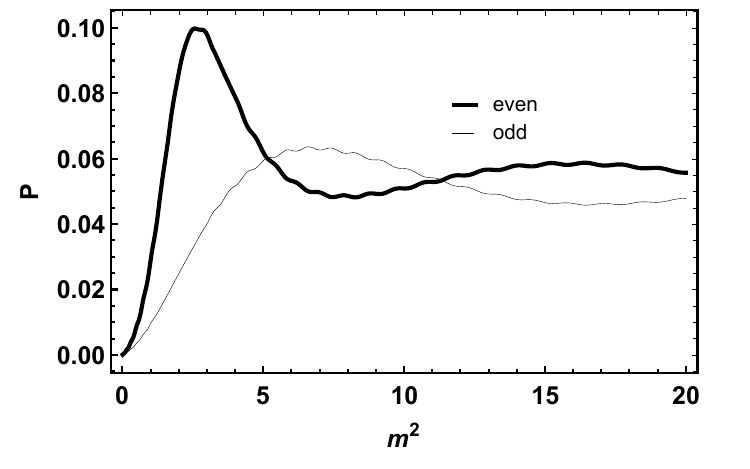}
\includegraphics[height=5cm]{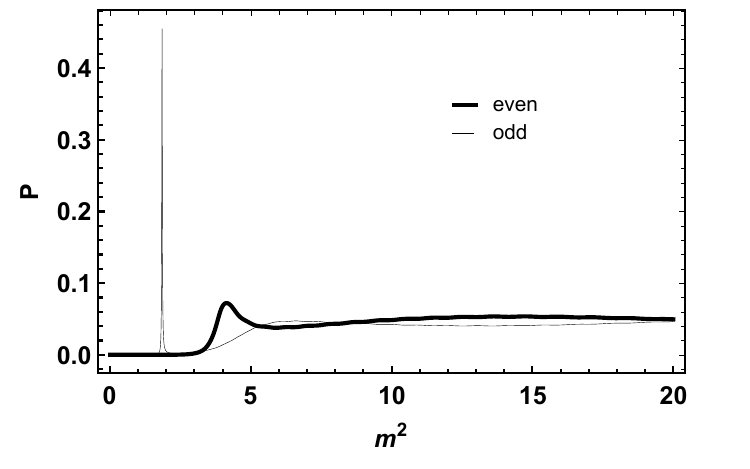}\\
(b) \hspace{6 cm}(c)
\end{tabular}
\end{center}
\caption{ Relative probability for $f_1$ with $\kappa_g=p=\lambda=1$. (a) $n=1$ and $k=0.1$. (b) $n=2$ and $k=0.01$. (c) $n=3$ and $k=0.001$.
\label{fig14}}
\end{figure}

In Fig.\ref{fig14}, the behavior of the relative probability for $f_1$ is depicted. Notably, for $n=1$, a probability peak gradually forms at $m^2=2.348$ for the odd solution (Fig.\ref{fig14}.a), indicating a potential resonant mode. With $n=2$, the probability peak emerges at $m^2=3.427$ for the even solution (Fig.\ref{fig14}.b). At $n=3$, the probability peaks become more distinct (Fig.\ref{fig14}.c). Specifically, resonant modes manifest at $m^2=4.104$ for the even solution and at $m^2=1.859$ for the odd solution.

\begin{figure}[ht!]
\begin{center}
\begin{tabular}{ccc}
\includegraphics[height=5cm]{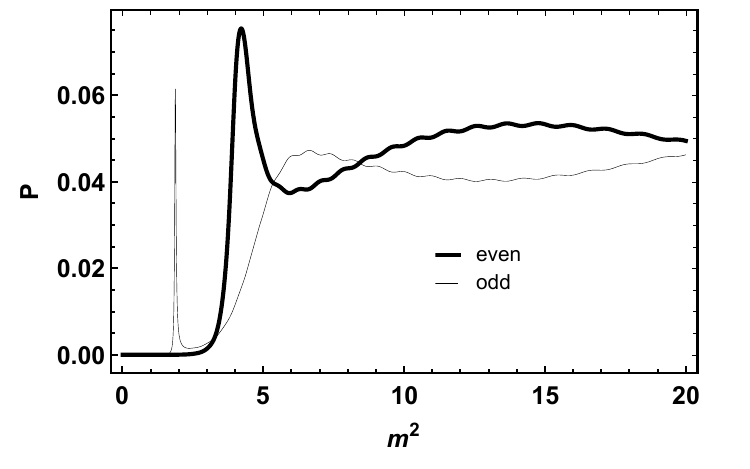}
\end{tabular}
\end{center}
\caption{ Relative probability for $f_2$ with $\kappa_g=p=\lambda=1$ and $k_{1,2}=0.001$.
\label{fig15}}
\end{figure}

Regarding $f_2$, the relative probability behavior is illustrated in Fig.\ref{fig15}. Notably, clear probability peaks are observed. For even solutions, the resonant mode emerges at $m^2=4.244$, while for odd solutions, it occurs at $m^2=1.963$.

\section{Final remarks}
\label{sec5}

In this study, we explored the impact of modifying the standard STEGR on fermion localization, utilizing two distinct $f(Q)$ models. Our investigations confirmed the physical viability of these models. We observed that variations in the parameters governing STEGR generalization lead to the emergence of new domain walls on the brane. This emergence triggers novel phase transitions in the scalar field, resulting in a double-kink solution. The brane's energy density responds to these structural formations, notably leading to brane splitting. To gain deeper insights into these newly formed structures, we analyzed brane stability through DCE. We found that the stability points of our models are intrinsically connected to brane splitting, representing the models' most probable configurations.

To facilitate the localization of fermions, we employed minimal Yukawa coupling between the fermion and the matter field. Our solutions for massless modes indicated strong localization on the brane, particularly for left-chirality fermionic modes. The fermions responded to the brane splitting, affecting both massless and massive modes. Notably, massive modes resembled free waves, suggesting these fermions likely escape the brane. However, near the brane core, massive modes exhibited intriguing behaviors, hinting at heightened sensitivity to gravitational changes.

For a nuanced analysis of fermion localization on the brane, we utilized two probabilistic measurement tools: Shannon entropy and relative probability. Shannon entropy revealed that deviations from standard STEGR decrease the certainty of fermion localization on the brane. These entropic measurements conformed to the BBM uncertainty relation, demonstrating a decrease in information loss (uncertainty in fermion locations) in position space and an increase in momentum space with amplified gravitational modifications. Additionally, relative probability enabled the identification of potential resonant modes in the model, indicating massive modes more likely to be localized on the brane.

Our findings underscore the effectiveness of probabilistic measurements (DCE, information measures, and relative probability) in examining the stability and identifying the most probable configurations of modified gravity models. Moreover, these tools prove adept at pinpointing configurations with a higher likelihood of fermion localization on the brane. Looking ahead, we aim to apply these methodologies to other modified gravity models, expanding our understanding of these complex systems.

\section*{Acknowledgments} One of the authors S.H. Dong would like to thank the partial support of project 20230316-SIP-IPN, Mexico. S.H. Dong started this work on the leave of IPN due to permission of research stay in China.

\end{document}